\documentstyle[12pt]{article}
\begin{document}

{\bf A Linear Quantum Dynamic Theory for Coherent Output of }

{\bf Bose-Einstain Condensation}

\bigskip

C. P. Sun$^{a,e}$, J. M. Li$^b$, H. Zhan$^c$, Y. X. Miao$^c$, S. R. Zhao$^d$
and G. Xu$^a$

$^a$Institute of Theoretical Physics, Academia Sinica, Beijing 100080, China

$^b$Department of Physics, Peking University, Beijing 100871, China

$^c$Institute of Theoretical Physics, Northeast Normal University, Changchun
130024, China

$^d$Department of Applied Mathematics, University of Western Ontario,
London, Ontario,

Canada, N6A, 5B7

$^e$Department of Physics, the Chinese University of Hong Kong, Shatin, NT,
Hong Kong

\bigskip\ 

\begin{center}
{\bf ABSTRACT}
\end{center}

\noindent A model for the coherent output coupler of the Bose-Einstein
condensed atoms from a trap in the recent MIT experiment (Phys. Rev. Lett., 
{\bf 78}(1997)582) is established with a simple many-boson system of two
states with linear coupling. Its exact solution for the many-body problem
shows a factorization of dynamical evolution process, i.e., the wave
function initially prepared in a direct product of a vacuum state and a
coherent state remains in a direct product of two coherent states at any
instance in the evolution of the total system. This conclusion always holds
even for a system with a finite average particle number in the initial
state. Its thermodynamical limit can be directly dealt with in the
Bogoliubov approximation and manifests that an ideal condensate in the trap
will remain in a coherent state after the r.f. interaction while the
output-coupler pulse of atoms is also in a coherent state, which means a
coherent output of atomic beam to form a macroscopic quantum state in a
propagating mode.

\bigskip 


\newpage\ 

\noindent {\bf 1. Introduction}

\noindent According to de Broglie matter wave theory, all matter in general
can behave itself like a light that spreads out in space and combines with
another one to form interference pattern. In principle, it is logical to
image that a ``matter wave laser'' can be realized, in analog to a light
laser, to produce the output coherent matter wave . Based on such
fundamental consideration and stimulated by the experiments observing atomic
Boson-Einstein condensate (BEC) in last two years [1-3], a number of efforts
coming from different research groups have been placed on both its
experimental setup and theoretical possibilities [4-7]. Until the end of
last year, a milestone experiment for the matter wave laser of bosonic atoms
(also called atom laser) was accomplished by Mewes et al in MIT [8].

In their experiment, they coupled the toms trapped in a state $|1\rangle $
with that untrapped in another state $|2\rangle $ by a sweeping radiation
frequency (r.f.) field. The r.f. field coherently turns the BEC atoms
constrained in $|1\rangle $ into a propagating mode of atoms in $|2\rangle $%
. The experiment shows that `` the condensate which is initially in a
coherent state remains in such a state after the r.f. interaction while the
output-coupler pulse of atom is also in a coherent state'' [8]. In fact, a
laser of photons above threshold can be understood in terms of the coherent
state for its well-defined classical analogy with minimum quantum
fluctuation in amplitude [9]. Similarly, it is reasonable to believe that
such a output of atoms in a single coherent state is something like the
photon laser. Roughly, we can regard the dynamic process of wave function
evolution driven by the r.f. field as a ``stimulated'' process of (massive)
matter wave.

In this paper, we first use the Bogoliubov approximation [10,11] in
thermodynamic limit to give a theoretical analysis for the dynamic lasering
process of BEC atoms in the MIT experiment. We will show that such an
approximation is not actually needed to reach the conclusion about coherent
output in the requirement of very large $N_c$. Namely, even for the case
that the number $N_c$ of atoms condensed in the ground state is not very
large, thereby the Bogoliubov approximation is violated, the output of
finite atoms is still coherent so long as the system is interatomic
interaction free and the atoms in the trap is initially in a coherent state.
Theoretically, this is the factorization structure for the evolution of the
many-boson model of two state with linear coupling, i.e., the wave function
at any instance remains in a direct product of two coherent states if the
system is initially prepared in a direct product of a vacuum state and a
coherent state. This fact does not depend on the average number of condensed
atoms in the initial state and it implies that the experiment result by
Mewes et al is quite steady even for rather finite $N_c$.\qquad \qquad

\noindent {\bf 2. A Dynamic Model of Output Coupler via Bogoliubov
Approximation}

\noindent The second quantization of model Hamiltonian for the MIT
experiment can be written as 
\begin{equation}
\label{1}H=\hbar \omega _ab_2^{\dag }b_2+\hbar \omega _R\left[ b_2^{\dag
}b_1\exp \left( -i\int_0^t\omega (\tau )d\tau \right) +h.c\right] 
\end{equation}
in terms of the creation and annihilation operators, $b_1^{\dag },$ $%
b_2^{\dag }$, $b_1$ and $b_2$, of bosonic atoms for the magnetically trapped
state $|1\rangle $ and the untrapped state $|2\rangle $ with level
difference ${\hbar }\omega _a$. The r.f. pulse of varying frequency $\omega
(t)$ is a sweeping classical electromagnetic field, which couples $|1\rangle 
$ and $|2\rangle $ through the dipole matrix element ${\hbar }\omega _R=%
\sqrt{\hbar \omega /{2\varepsilon _0V}}\equiv {\hbar }g/\sqrt{V}$. $V$ is
the effective mode volume and $\varepsilon _0$ is the vacuum permittivity.
This many atom system without interatomic interaction can be modeled as a
linear coupling system of two-oscillators. The main simplification is to
ignore the quantized motion of atomic center of mass in the trapped state by
an inhomogeneous magnetic field. As pointed out by Mewes et al, this
inhomogeneous feature can be neglected for a sufficiently short r.f. pulse.

The MIT experiment is theoretically described as an initial value problem
for Sch\"ordinger equation governed by $H$ with the initial state $|\psi
(0)\rangle =|\alpha =\sqrt{N_c}\rangle _1\otimes |0\rangle _2$. Here, $%
|\alpha =\sqrt{N_c}\rangle _1$ is a Glauber coherent state of the operator $%
b_1$ characterizing $N_c$ atoms condensed in the trapped state $|1\rangle $.
No atoms occupy the untrapped state $|2\rangle $ initially. Why does the
coherent state $|\alpha \rangle $ represent the BEC is still an open
question, but there are some reasons enable one to believe it is right, such
as the phase locking implied by BEC [12] and the correct average atomic
number $N_c$. The additional reason we believe is due to the connection with
the Bogoliubov approximation [10,11], which replaces operators $b_1$ and $%
b_1^{\dag }$ with a $c$-number $\sqrt{N_c}$ when much many atoms condensate
in the trapped state. A factorizable evolution structure in the MIT
experiment is just rooted in this connection and will be discussed as
following. There is a close relation between the selection of initial
coherent state $|\alpha \rangle $ and the Bogoliubov approximation. In fact,
a unitary transformation $|\phi (t)\rangle =D^{-1}(\alpha )|\psi (t)\rangle $%
, in terms of the generator $D(\alpha )=\exp [\alpha b_1^{\dag }-\alpha b_1]$
of coherent state, defines an equivalent initial value problem of
Sch\"ordinger equation for $|\phi (0)\rangle =|0\rangle _1\otimes |0\rangle
_2$ and the equivalent Hamiltonian%
$$
D(\alpha )^{-1}HD(\alpha )=H+\hbar \omega _R\sqrt{N_c}\left[ b_2^{\dag }\exp
\left( -i\int_0^t\omega (\tau )d\tau \right) +h.c\right] . 
$$
For the case of very large $N_c$ in BEC, the coupling term is very small in
comparison with the term of ${\hbar }\omega _R\sqrt{N_c}$ and can be
neglected to obtain an effective Hamiltonian 
\begin{equation}
\label{2}{\cal H}_B=\hbar \omega _ab_2^{\dag }b_2+\hbar \omega _R\sqrt{N_c}%
\left[ b_2^{\dag }\exp \left( -i\int_0^t\omega (\tau )d\tau \right)
+h.c\right] .
\end{equation}
This is just the Bogoliubov approximation of the original Hamiltonian $H$.

Now let's consider the thermodynamical limit of an infinite number of atoms
in an infinite volume but with the density fixed, i.e., $n_c=\lim
_{N,V\rightarrow \infty }\left( N_c/V\right) \rightarrow constant$, the
coupling term $\hbar \omega _R\propto {1/}\sqrt{V}\rightarrow 0$, but ${%
\hbar }\omega _R\sqrt{N_c}=\hbar g\sqrt{n_c}$ is finite. Therefore, in the
point of view of Sch\"ordinger evolution, the Bogoliubov approximation for
our problem just describes an excitation motion of system in a large
background provided by the BEC initial state. Namely, for a very large $N_c$%
, the system initially in a coherent state remains in such a state while
another component $|\Phi (t)\rangle $ is governed by the Bogoliubov
approximate Hamiltonian. Mathematically, the evolution wave function enjoys
a factorizable structure 
\begin{equation}
\label{3}|\psi (t)\rangle =|\alpha \rangle \otimes |\Phi (t)\rangle . 
\end{equation}
\qquad

The similar factorization structure has been shown in the analysis about the
transition from quantum domain to macroscopic, classical domain concerning
the wave function collapse in quantum measurement [13]and the macroscopic
damping in quantum dissipation [14,15]. The method calculating wave function
developed in [14,15] will be used in this paper to obtain the exact solution
for factorized wave function in section 4.\qquad 

\noindent {\bf 3. Coherent Output via Factorization}

\noindent The next step is to prove $|\Phi (t)\rangle $ is still a coherent
state evolving from its vacuum state $|0\rangle _2$. It is quite clear that
the Bogoliubov approximation Hamiltonian (2) is just a forced harmonic
oscillator (FHO). It is well-known that the displacement property of a
quantized FHO results dynamically in a coherent state for the initial vacuum
state. By considering the case of $\omega (\tau )$ independent of time,
i.e., $\omega (\tau )=\omega $, the solution of the Hamiltonian (2) can be
found in some standard textbook [16] $|\Phi (t)\rangle =|\tilde \alpha
(t)\rangle $ with 
\begin{equation}
\label{4}\tilde \alpha (t)=\frac{\omega _R\sqrt{N_c}}\Delta \left(
e^{i\Delta t}-1\right) e^{-i\omega _at},
\end{equation}
Namely, the total wave function in the BEC with large $N_c$ can be
factorized in the direct product of two coherent states $|\psi (t)\rangle
=|\alpha \rangle \otimes |\tilde \alpha (t)\rangle $. The second components
implies the output-couple pulse of the atomic beam is in the coherent state,
which ``can be regarded as a pulsed atomic laser''.

Notice again that the above theoretical discussion about the output- coupler
for atomic laser is only with the help of a highly-simplified many- boson
model of two states with linear coupling rather than the complicated
nonlinear one, such as the system of Gross-Petaevskii equation. Because the
atomic number of the coherent output in the untrapped state 
\begin{equation}
\label{5}\langle \psi (t)|b_2^{\dag }b_2|\psi (t)\rangle =|\tilde \alpha
|^2=2\frac{\omega _R^2}{\Delta ^2}N_c\left( 1-\cos \Delta t\right) 
\end{equation}
is an oscillating function of $t$, the sweeping mechanism [17, 8] of the
r.f. pulse must be invoked here to localized the population in the untrapped
state at $t=\infty $. This will be in the aid of Landau-Zener method [17] to
control the populations in trapped and untrapped states by adjusting the
time- dependent frequency $\omega (t)$ of the r.f. pulse from diabetic to
adiabatic point in the avoid level crossing. A direct treatment for the r.f.
sweeping is changing the factors $\exp \left( \pm i\omega t\right) $ into $%
\exp \left( \pm i\int_0^t\omega (\tau )d\tau \right) $ of $H$ in the above
approach. Near the resonance $\omega (t_0)=\omega _a,$ $\bigtriangleup
\simeq \dot \omega (t_0)(t-t_0)$, 
\begin{equation}
\label{6}\tilde \alpha (\infty )=ig\sqrt{n_c}\int_0^\infty \exp \left( \frac
i2\dot \omega (t_0)(t-t_0)^2\right) dt=ig\sqrt{\frac{\pi n_c}{4\dot \omega
(t_0{)}}}. 
\end{equation}
This leads to a fixed coherent population in the untrapped state, which can
be controlled by adjustment of the changing rate $\dot \omega (t_0)$ of the
frequency of the r.f. pulse at resonance.\qquad

\noindent {\bf 4. Coherence From Finite Condensed Atoms}

\noindent The above discussion shows that the factorized structure of wave
function crucially result in the output of coherent atomic beam as an atomic
laser when the initial state of system is prepared on the BEC in a trap for
which the Bogoliubov approximation is hold. This fact seems that the
factorizable structure of evolution depends on the Bogoliubov approximation,
namely, it is decided by whether or not there are very many atoms in the BEC
trapped state. In the following, it will be demonstrated by the exact
solution of the Hamiltonian (1) that the requirement of very large $N_c$ is
not necessary so long as the atoms are prepared in a coherent state $|\alpha
\rangle $ even with rather finite average atom number${\rm ~}N_c$ $=$ $%
\langle \alpha |b_2^{\dag }b_2|\alpha \rangle .$ The coherent state is only
a special Poisson superposition of infinite number states with marched
phases $\Theta _n=n\theta ,$ $n=1,2,\ldots .$

The second reason to consider the exact solution concerns the analogy in the
photon case. In the theory of light laser, a sufficient strong light field
can be treated as a classical electromagnetic field and it can be regarded
as a coherent state above the threshold [9]. In this sense, the quantum
effect mainly results from the quantum fluctuation, which is demonstrated by
a weak light field. Therefore, for the matter wave of massive particles, the
infinite number case only corresponds to the classical analogy of photon
field. Then, we need to study the case corresponding to the quantum
fluctuation of the weak light field, for which the Bogoliubov approximation
is broken down.

According to [14,15], the exact solution of the wave function of the linear
system in the Schr\"odinger picture can be easily obtained from that of the
canonical operators, such as $b_1(t)$ and $b_2(t),$ in Heisenberg picture.
In the case of $\omega (\tau )$ independent of time, by solving a system of
one-order differential equations resulting from the Heisenberg equations of $%
b_1(t)$ and $b_2(t),$ the exact expressions of these solutions are 
\begin{equation}
\label{7}b_1(t)=\alpha _1(t)b_1(0)+\alpha _2(t)b_2(0),
\end{equation}
\begin{equation}
\label{8}b_2(t)=\beta _1(t)b_1(0)+\beta _2(t)b_2(0)
\end{equation}
where 
\begin{equation}
\label{9}\beta _1(t)={\frac{-\omega _R}{\sqrt{\Delta ^2+4\omega _R^2}}}%
\left[ e^{i\omega _{+}t}-e^{i\omega _{-}t}\right] e^{-i\omega t},
\end{equation}

\begin{equation}
\label{10}\beta _2(t)=\frac 1{\sqrt{\Delta ^2+4\omega _R^2}}\left[ \omega
_{+}e^{i\omega _{+}t}-\omega _{-}e^{i\omega _{-}t}\right] e^{-i\omega t}, 
\end{equation}
\begin{equation}
\label{11}\alpha _1(t)=\frac 1{\sqrt{\Delta ^2+4\omega _R^2}}\left[ \omega
_{+}e^{i\omega _{-}t}-\omega _{-}e^{i\omega _{+}t}\right] , 
\end{equation}
\begin{equation}
\label{12}\alpha _2(t)={\frac{-\omega _R}{\sqrt{\Delta ^2+4\omega _R^2}}}%
\left[ e^{i\omega _{+}t}-e^{i\omega _{-}t}\right] , 
\end{equation}
\begin{equation}
\label{13}\omega _{\pm }=\frac 12\left( \Delta {\pm }\sqrt{\Delta ^2+4{%
\omega }_R^2}\right) . 
\end{equation}
From the property of the evolution operator, $U(t)$, of the total
system,i.e., $b_\alpha (t)=U(t)^{\dag }b_\alpha (0)U(t)$, $(\alpha =1,2)$,
the initial state $|\psi (0)\rangle =|\alpha =\sqrt{N_c}\rangle \otimes
|0\rangle $ will evolve into 
$$
|\psi (t)\rangle =U(t)|\psi (0)\rangle =e^{-\frac 12|{\alpha }%
|^2}\sum_{n=0}^\infty {\frac{\alpha ^n}{n!}}b_1^{\dag n}(-t)|0\rangle
\otimes |0\rangle 
$$
\begin{equation}
\label{14}=e^{-\frac 12|{\alpha }|^2}\sum_{n=0}^\infty {\frac{\alpha ^n}{n!}}%
\left( \alpha _1(-t)b_1^{\dag }(0)+\alpha _2(-t)b_2^{\dag }(0)\right)
^n|0\rangle \otimes |0\rangle . 
\end{equation}
That is just a direct product of two coherent states 
$$
|\psi (t)\rangle =|\alpha \alpha _1(-t)\rangle \otimes |\alpha \alpha
_2(-t)\rangle 
$$
\begin{equation}
\label{15}=|\sqrt{N_c}e^{-\frac i2\Delta t}\left( \cos (\Omega t)+i\sin
(\Omega t)\cos \theta \right) \rangle \otimes |-i\sqrt{N_c}e^{-\frac
i2\Delta t}\sin \theta \sin (\Omega t)e^{-i\omega t}\rangle 
\end{equation}
where $tg\theta =2\omega _R{/}\Delta $, where $\Omega =\sqrt{\Delta
^2/4+\omega _R^2}.$

This result indeed leads to a coherent output of atoms in the untrapped
state so long as the atoms are initially prepared as the BEC in trapped
state even for the finite average number of atoms. In this output state with
wave vector $k$, the expectation value 
\begin{equation}
\label{16}\langle \hat \phi (x)\rangle =\langle \psi (t)|\hat \phi (x)|\psi
(t)\rangle =\sum_k\sqrt{\frac{2\hbar n_c}{{\omega }}}\sin \theta \sin
(\Omega t)\sin \left( kx-\frac 12\Delta t-\omega t\right) 
\end{equation}
of the bosonic field, $\hat \phi (x)=\sum_k\sqrt{\hbar /2\omega V}\left(
b_{2k}e^{ikx}+h.c\right) $, looks just like a classical coherent field,
where the variance of the field $\Delta \phi =\sqrt{\langle \hat \phi
(x)^2\rangle -{\langle \hat \phi (x)\rangle }^2}$ do not depend on $N_c$.
This means the relative uncertainty of the amplitude, $\Delta \phi /{\langle
\hat \phi (x)\rangle }$, will approach zero only in the thermodynamics
limit: $N_c,$ $V\rightarrow \infty ,$ but $N_c/V\rightarrow constant.$
Therefore, for finite $N_c$, the quantum fluctuation $\Delta \phi \propto 
\sqrt{h}$ can be neglected for the dynamic problem. The Bogoliubov
approximation in the last section is just the thermodynamics limit of the
results in this section. Notice that, in the thermodynamics limit, we have $%
\omega _R\rightarrow 0,$ $\Omega \rightarrow \frac \Delta 2$, but $N_c\sin
\theta \propto g\sqrt{n_c}$ do not approach zero.

\qquad

\noindent {\bf 5. Discussions}

\noindent In summary, for the MIT experiment on the output coupler of atomic
laser, we have firstly shown that, if the atoms with an average number $N_c$
are accumulated in a single quantum state $|1\rangle $ to form a coherent
state $|\alpha =\sqrt{N_c}\rangle _1$ and no atoms occupy the untrapped
state $|2\rangle $, the spin-like dynamics of Rabi oscillation will create a
factorized wave function $|\psi (t)\rangle =|\alpha (t)\rangle _1\otimes
|\beta (t)\rangle _2,$ which is a product of two coherent states. Even for
the case without ideal Bose-Einstein condensation that $N_c$ is not
sufficiently large and thus $|\alpha =\sqrt{N_c}\rangle _1$ is not a
macroscopic quantum state, this wave-function of evolution remains in a
factorizable structure so long as $|\alpha =\sqrt{N_c}\rangle _1$ is
coherent states. However, it seems very difficult to prepare finite numbers
of atoms in a coherent state duo to the massive feature of atoms. The
spin-like dynamics of Rabi oscillation will only create an entanglement
evolution state $|\psi (t)\rangle $ $=$ $\sum_{i,j}|\alpha _j(t)\rangle
_1\otimes |\beta _i(t)\rangle _2$ for an arbitrary initial components $|\psi
(0)\rangle _1$ other than a coherent state. Only in BEC case that the fixed
phase and amplitude of macroscopic wave function are defined and thus to
form a Glauber coherent state, this entanglement state can be completely
factorized.

Such a factorized structure of wave function first pointed out by Mewes et
al is very crucial to realize their experiment, but it dynamical origins
mainly depend on the linearity of coupling. For the case with an interatomic
interaction in the trapped or untrapped states, which roughly is of the form 
$b_i^{\dag }b_i^{\dag }b_ib_i,$ the corresponding Heisenberg equations are
no longer linear and thus this factorized structure will be broken down.
Therefore, the sufficiently weak interaction should be required for a
coherent output in practical system. In fact, Ballagh et al [18] have
modeled a theory for the evolution of two component BEC by generalizing the
Gross-Pitaevskii equation. Their theory introduces a non-linear element
resulting from the interatomic interaction and shows a rich variety of
phenomena of nonlinear effect such as the localization of atom in the
untrapped state due to the nonlinear coherent coupling. 

Notice that an interference experiment was also carried out by Andrews et al
in the same research group at MIT [19] to examine the coherence of atoms in
BEC by overlapping two expanding condensates flying from the double well
magnetically-trapped potential when this trap is adiabatically switched off.
The next step following this experiment was immediately accomplished to test
the coherence of the output couplers rather than a pair of expanding atomic
sources. They also have shown that a similar high-contrast regular pattern
of interference. Based on an intuitional understanding of the MIT ``atomic
laser'' in terms of coherent states described in this paper, a many--body
quantum theory for the interference of output couplers can be present in a
forecoming paper as a space-dependent generalization of present results. The
non-vanishing order parameter of quantum system, concerning a symmetry
broken involving the atomic number conservation, can be obtained for the
description of regular pattern of stripes in these interference experiments.

\bigskip

One (CPS) of t he authors wish to express his sincere thanks to K. Young for
inviting him to visit the Chinese University of Hong Kong as a C. N. Yang's
Fellow. The work is supported in part by the NSF of China.

\newpage 

{\bf References}

\begin{enumerate}
\begin{description}
\item[1]  M. H. Anderson, J. R. Ensher, M. R. Matthews, C. E. Wieman and
E.A. Cornell, {\it Science}, {\bf 269}(1995)198.

\item[2]  C.C.  Bradley, C. A. Sackett, J. J. Tollett and R. G. Hulet, {\it %
Phys. Rev. Lett}., {\bf 75}(1995)1687.

\item[3]  K. B. Davis, M. -O. Mewes, M. R. Andrews, N. J. van Druten, D.
S.Durfee, D. M. Kurn and W. Ketterle, {\it Phys. Rev. Lett.}, {\bf 75}
(1995)3969.

\item[4]  .H.Wiseman and M.Collett, Phys.Lett.A.,202(1995),246

\item[5]  R.Spreeuw,T.Pfau, U.Janicke,and M.Wilkens,{\it \ Europhys.Lett}%
.,32(1995),469

\item[6]  M.Holland et al.{\it , Phys.Rev.A}{\bf .,} 54(1996), R1757.

\item[7]  A.Guzman, M.Moore,and P.Meystre, Phys.Rev.A53(1996), 977

\item[8]  M. Mewes, M. R. Andrews, D. S. Durfee, D. M. Kurn, C. G. Townsend
and W. Ketterle, {\it Phys. Rev. Lett}., {\bf 78}(1997)582.

\item[9]  R.Loudon,{\it The Quantum Theory of Light, Oxford Univ.Press,1983,
Chapters 4 and 7}

\item[10]  N. N. Bogoliubov,{\it \ J. Phys. }(USSR), {\bf 11}(1947)23.

\item[11]  A. L. Fetter and J. D. Walecka, {\it Quantum Theory of
Many-Particle System}, McGraw-Hill, 1971
\end{description}

\begin{description}
\item[12]  P. Nozi\`eres, in {\it Bose-Einstein Condensation}, A. Griffin et
al (eds), Cambridge Univ. Press, 1995, pp. 15-30.

\item[13]  C.P.Sun, {\it Phys. Rev}., {\bf A48}(1994),898

\item[14]  L. H. Yu and C. P. Sun, {\it Phys. Rev}., {\bf A49}(1994)592.

\item[15]  C. P. Sun and L. H. Yu, {\it Phys. Rev.,} {\bf A51}(1995)1845.

\item[16]  W. H. Louisell, {\it Quantum Statistical Properties of Radiation}%
, John Wiley \& Sons, New York, 1973, pp. 203-207.

\item[17]  J. Rubbmark, M. M. Kash, M. G. Littman and D. Kleppner, {\it %
Phys. Rev.,}{\bf \ A23}(1981)3107.

\item[18]  R. J. Ballagh, K. Burnett and T. F. Scott, {\it Phys. Rev. Lett.}%
, {\bf 78}(1997)1607.

\item[19]  M. R. Andrews, C. G. Townsend, H. J. Miesner, D. S. Durfee, D.
M.Kurn and W. Ketterle, {\it Science}, {\bf 275}(1997)637.
\end{description}
\end{enumerate}

\end{document}